\documentclass[twocolumn,preprintnumbers,amsmath,floatfix,prl,aps,superscriptaddress]{revtex4-1}

\usepackage{graphicx}% Include figure files
\makeatletter

\def\V{{V_{\rm pipe}}}

\begin{document} 

\title{Heat pumping in nanomechanical systems}

\author{Claudio Chamon} \affiliation{Department of Physics, Boston
  University, Boston, Massachusetts 02215, USA}

\author{Eduardo R. Mucciolo} \affiliation{Department of Physics,
  University of Central Florida, Orlando, Florida 32816, USA}

\author{Liliana Arrachea} \affiliation{Departamento de F\a'{i}sica,
  FCEyN and IFIBA, Universidad de Buenos Aires, Pabell\'on 1, Ciudad
  Universitaria, 1428 Buenos Aires, Argentina}

\author{Rodrigo B. Capaz} \affiliation{Instituto de F\a'{i}sica,
  Universidade Federal do Rio de Janeiro, 
%C. P. 68528, 
Rio de Janeiro
  21941-972, RJ, Brazil}

\date{\today}

%%%%%%%%%%%%%%%%%%%%%%%%%%%%%%%%%%%%%%%%%%%%%%%%%%%%%%%%%%%%%%%%%%%%
\begin{abstract}

%We construct a quantum  mechanical analog for phonons of the Bartoli
%cycle used by Boltzmann to obtain light pressure in blackbody
%radiation. This quantum phonon pump can coherently transfer heat from
%a cold to a hot body connected by a nanomechanical system.

  We propose using a phonon pumping mechanism to transfer heat from a
  cold to a hot body using a propagating modulation of the medium
  connecting the two bodies. This phonon pump can cool nanomechanical
  systems without the need for active feedback. 
  % We start by analyzing the thermodynamics of a simplified case of a
  % cycle where one of the strokes corresponds to a traveling fully
  % reflective barrier, and then consider the effect of a
  % semireflective barrier.
  We compute the lowest temperature that this refrigerator can
  achieve.
  % , which depends on the hot reservoir temperature, the reflectivity
  % of the barrier, the ratio of the barrier speed to the phonon
  % velocity, and the lattice anharmonicity.

\end{abstract}
%%%%%%%%%%%%%%%%%%%%%%%%%%%%%%%%%%%%%%%%%%%%%%%%%%%%%%%%%%%%%%%%%%%%%

\pacs{}

\maketitle

%%%%%%%%%%%%%%%%%%%%%%%%%%%%%%%%%%%%%%%%%%%%%%%%%%%%%%%%%%%%%%%%%%%%%%%%
%%%%%%
% Introduction

Freezing out atomic motion by cooling matter to absolute zero
temperature is a thought that has, for ages, fascinated both
scientists and laymen alike. In atomic gases, techniques such as
evaporative cooling can bring temperatures down to the submicrokelvin
scale, allowing for the observation of quantum phenomena such as
Bose-Einstein condensation. In solid state matter, the ionic motion
takes the form of oscillations around equilibrium positions, and
completely freezing the system (in the case of an insulator) means
removing all lattice vibrations -- phonons -- leaving solely the
quantum mechanical zero-point motion.

The quest for observing quantized mechanical motion in macroscopic
systems has incited several experimental groups in recent years
\cite{sciencereview}. In most cases, cooling is obtained by a feedback
mechanism which involves optical or electronic sensors and some
control system that acts directly on a cantilever.  In this Letter, we
argue that it is possible to cool a nanomechanical system without
relying on feedback control. The mechanism we propose acts directly on
the acoustic phonons carrying heat in and out of the system without
the need for monitoring its state. By deforming the lattice in the
medium connecting the mechanical system to its phonon thermal
reservoir, one can pump heat against a temperature gradient by
extracting out phonons. The mechanism resembles a classical cooling
cycle of a thermal machine and its physical basis is time-reversal
symmetry breaking. The pump works in both coherent and incoherent
phonon regimes.

Quantum coherent electron pumps have been studied extensively since
Thouless's original proposal \cite{thouless83}. For instance, using
lateral quantum dots and quantum wires, charge \cite {brouwer98}, spin
\cite{sharma01}, and heat \cite{arrachea05} currents can be created in
the absence of bias by modulating adiabatically and periodically in
time two independent external parameters. In contrast, pumping
massless bosons such as acoustic phonons is a much more subtle
problem. For one, it is much harder to pump adiabatically phonons due
to the lack of a large energy scale such as the Fermi energy.
%The need to integrate over all frequencies renders the adiabatic
%regime untenable.
Moreover, phonons not only obey a different wave equation but are also
not conserved when scattered by external perturbations that couple
linearly to the displacement field ({\it i.e.}, a driving
force). The result in this case is entropy generation in addition to
pumping.

In practice on can pump phonons with minimum heat generation by
coupling quadratically to the displacement field, either by locally
modulating the propagation velocity or by locally applying a pinning
potential. An extreme example of a pinning perturbation, which
preserves phonon number, is one that imposes Dirichlet boundary
conditions to the displacement field at a given point in space. When
such a perturbation travels along a quasi-one-dimensional medium, it
works as a linear peristaltic pump. Below, we show that this mechanism
allows for cooling down the system to a minimum temperature $T_{\rm
  min}$ which, in one-dimension, is given by the expression
\begin{equation}
\label{eq:Tmin}
T_{\rm min} = \sqrt{\sqrt{ \Theta_B^4 + T_H^4} - \Theta_B^2},
\end{equation}
with $\Theta_B = \lambda \sqrt{5v_B/2\pi^3 c}$, where $T_H$ is the
temperature of the hot thermal reservoir, $\lambda$ is the
perturbation strength, $c$ is the phonon velocity, and $v_B$ is the
barrier speed.

\begin{figure}[htp]
\centering
\includegraphics[width=6cm]{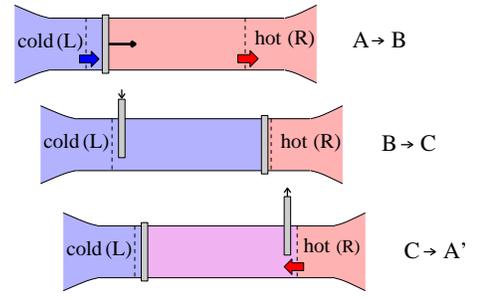}
\caption{Pumping cycle:
  A$\rightarrow$B$\rightarrow$C$\rightarrow$A$^\prime$ (see text for
  an explanation). A traveling lattice perturbation acts as a
  semireflective barrier moving from cold to hot reservoir. The wide
  arrows indicate unimpeded heat flow.}
\label{fig:cycle}
\end{figure}

A scheme of the pumping cycle is shown in Fig. \ref{fig:cycle}, where
the nanomechanical system to be cooled is represented by the left
(cold) side. The local modulation in the phonon velocity or pinning
potential works like a moving semireflective barrier to the
phonons. In process A$\to$B, the barrier is translated from the cold
to the hot side of a cavitylike region. After it reaches the endpoint,
another barrierlike perturbation is activated at the opposite side of
the cavity (process B$\to$C). Then, in C$\to$A$^\prime$, the first
barrier is deactivated and phonons from the hot reservoir free expand
into the cavity. The procedure is then repeated.

%In the barrier's reference frame, there is a Doppler shift in
%phonon frequency which depends on the direction of propagation: those
%phonons moving against the barrier have their frequency increased,
%while the opposite occurs with those phonons moving in the same
%direction as the barrier. Since the barrier reflectivity is frequency
%dependent, its motion creates an imbalance between heat fluxes.

Interesting issues arise out of this simple process of moving a
reflective barrier (a ``mirror'') for phonons, in particular that of
phonon pressure across the barrier. Indeed, a similar process to the
one described above was used by Bartoli when he attempted to show the
applicability of thermodynamics to electromagnetism and raised the
question of radiation pressure~\cite{bartoli}, which in turn inspired
Boltzmann in his studies of blackbody radiation~\cite{boltzmann}. The
issue of phonon pressure is not trivial (and more subtle than the case
of photons) as phonons carry crystal momentum (${\bf q}$) but not
obviously physical linear momentum (denoted by ${\bf p}$). The
connection between these two forms of momentum requires anharmonicity
and is given by ${\bf p}_{\bf q} = \gamma\, d \hbar {\bf q}$, where
$\gamma$ is the Gr\"uneisen parameter of the lattice and $d$ denotes
the spatial dimension \cite{AshcroftMermin,sorbello,lee}. This impacts
the relation between pressure and energy density in a phonon gas; for
instance, in the case of a single acoustic mode, the relation takes
the simple form $p = -\left( \partial F/\partial V\right)_T = \gamma
E/V$.

We begin by discussing first the case of a fully reflective
barrier. We can treat the problem as a gas of phonons, which we cycle
according to Fig.~\ref{fig:cycle}. Notice that the barrier does not
let heat pass through and the cooling is due to the removal of
internal energy from the left-hand side, dumping it into the
right-hand side, as explained below.

The expansion $A\to B$ is adiabatic and reversible ($\Delta
S_{R,L}=0$, i.e., no heat exchange between left- and right-hand
sides). Recalling the standard equation for massless bosons, $dS =
dE/T + E\, dV/VTd$, we can relate changes in energy to variations in
volume. When the barrier moves to the right, the change in internal
energies on the two sides are $E_L^B = E_L^A - p_L \V/\gamma d$ and
$E_R^B = E_R^A + p_R \V/\gamma d $, where $\V$ is the swept
volume. Then, once we insert the other barrier to get to $C$, we
redraw the boundary of what $L$ is. The volume of $L$ changes by a
factor $(V_L-\V)/V_L$. So in $C$ we have $E_L^C =
\left(1-\frac{\V}{V_L}\right) E_L^B$, $E_R^C=E_R^B$, and $E_{\rm
  pipe}^C = \frac{\V}{V_L}\;E_L^B$, where the last energy is the one
inside the ``pipe''. Then, once the right barrier is removed in going
$C\to A'$, one redraws the boundary of what $R$ is, so
$E_L^{A'}=E_L^C$ and $E_R^{A'}=E_R^C+E_{\rm pipe}^C$.
\begin{subequations}
Putting it all together, we have
\begin{eqnarray}
\label{eq:eL}
\Delta E_L^{A\to A'} & = & \left(1-\frac{\V}{V_L}\right) \left(E_L^A
-\frac{p_L}{\gamma\, d} \V \right) - E_L^{A} \nonumber\\ & = &
-\left(e_L+\frac{p_L}{\gamma\, d} \right) \V + \dots,\\
%\end{eqnarray}
%
%\begin{eqnarray}
\label{eq:eR}
\Delta E_R^{A\to A'} & = & \frac{\V}{V_L}\;\left(E_L^A
-\frac{p_L}{\gamma\, d} \V \right) + p_R \V \nonumber\\ & = &
  \left(e_L+\frac{p_R}{\gamma\, d} \right) \V + \dots,
\end{eqnarray}
where $\dots$ stand for terms down by powers of $\V/V_{L,R}$, and
$e_{R,L}=E_{R,L}/V_{R,L}$ are the intensive energy densities in the
two sides. All the work done occurs in $A\to B$ and is given by
\begin{equation}
\label{eq:W}
W^{A\to A'} = \frac{(p_R-p_L)}{\gamma\, d}\,\V + \dots,
\end{equation}
\end{subequations}
where the leading term is insensitive to changes in the pressures
$p_{L,R}$ as the volume expands. The unusual relation between work and
volume change shown in Eq. (\ref{eq:W}) comes from the fact that, in
our scheme, volume changes also require an increase in the number of
unit cells, so that the lattice unit cell volume is kept constant
(i.e., no compression). The work required to add units cells leads to
the $\gamma\, d$ factor dividing the pressure difference. (Notice that
this factor is absent for photons, since $\gamma =1/d$ follows from
${\bf p}_{\bf q} = \hbar {\bf q}$. In the case of light, there is no
undelying lattice system -- an ``ether'' -- that needs to be accounted
for.) In the process $A\to A'$ described above, all entropy increase
occurs when the barrier is removed in going $C\to A'$, and the second
law of thermodynamics is satisfied. From this analysis, we can compute
the energy flux out of the left reservoir per unit time of operation
of the cycle:
\begin{equation}
\label{eq:J-gas}
{\cal J}_L^E=\left(e_L+\frac{p_L}{\gamma\, d}\right)\, v_B,
\end{equation}
where $v_B$ is the barrier speed. Here we use for total time the
duration of the $A\to B$ stroke, assuming that the equilibration in
the entropy production part $C\to A'$ is fast compared to this time.

%%%%%%%%%%%%%%%%%%%%%%%%%%%%%%%%%
%%%%%%%%%%%%%%%%%%%%%%%%%%%%%%%%%

For our case of interest, $e_L= \eta_d T_L^{d+1}/c^d$, where $\eta_d =
2g\, d!\, \zeta(d+1)/[(4\pi)^{d/2}\Gamma(d/2)]$ with $\zeta(z)$ and
$\Gamma(z)$ denoting the Riemann zeta and Gamma functions,
respectively, while $g$ is a degeneracy factor. Notice that the energy
flux depends only on the intensive quantities for the system on the
left (and thus on $T_L$), and not on any property on the right-hand
side of the barrier, in particular its temperature. This is a
straightforward consequence of the fact that the barrier is perfectly
reflective, so one is not faced with the difficulty of fighting a
thermal gradient between the hot and cold reservoirs. The idealized
situation, however, serves the purpose of displaying clearly the main
principle of our cooling mechanism.

Let us turn the discussion to the less idealized situation when the
barrier is not perfectly reflective, allowing some heat to be
transmitted from the hot to the cold side. In this case, we
intuitively expect that the slower we move the barrier, the more
difficult it becomes to cool, because the energy transferred in the
operation $A\to B\to C\to A'$ depends only on the volume swept by the
barrier, but not on the rate (as long as the $A\to B$ stroke is done
in a quasi-equilibrium situation, allowing for thermal equilibration
on both sides of the barrier). In addition, the longer we take to move
the barrier to the right in the $A\to B$ stroke, the more heat is
transferred through the transmitting barrier (the total transfer
scales linearly with the sweeping time). So let us now compute the
heat flow through the moving barrier, and the conditions to attain net
cooling for a semireflective barrier moving with speed $v_B$.
Hereafter, for simplicity, we focus on a purely one-dimensional case
($d=1$).

For concreteness, consider a ``moving mirror'' corresponding to a
region in space where the atoms are coupled to an external short-range
potential, which is localized in space. The position of this pinning
potential is modulated in time so as to make it travel at speed $v_B$,
causing the reflection and transmission coefficients to depend on the
red and blue shifted frequencies of the phonons coming from the two
reservoirs. Acoustic phonons in a one-dimensional chain, interacting
with such a ``moving mirror'' potential of strength $\lambda$, obey
the following wave equation in the continuum limit:
\begin{equation}
\partial_t^2 u(x,t) - c^2 \partial_x^2 u(x,t) = -\lambda\, c\,
\delta(x-v_B\, t)\;u(x,t),
\label{eq:lab1}
\end{equation}
It is simpler to work in the
reference frame of the barrier, $t^\prime = t$ and $x^\prime = x - v_B
t$, where the wave equation becomes
\begin{equation}
\label{eq:mirrorweq}
\left[ \left( \partial_{t^\prime} - v_B\, \partial_{x^\prime}
  \right)^2 - c^2 \partial_{x^\prime}^2 \right] u(x^\prime,t^\prime) =
-\lambda\, c\, \delta(x')\, u(x',t').
\end{equation}

Let us consider plane wave solutions to Eq.~(\ref{eq:mirrorweq}) in
the two regions, to the left of the barrier (with amplitudes
$A_\omega^-$ and $B_\omega^-$) and to its right (with $A_\omega^+$ and
$B_\omega^+$):
\begin{equation}
\label{eq:mirror-frame-sol}
u_\pm(x^\prime,t^\prime) = \int\! d\omega\, e^{i\omega t^\prime}
\left( A^\pm_\omega e^{-i\omega x^\prime /v_R} + B^\pm_\omega
e^{i\omega x^\prime /v_L} \right),
\end{equation}
with $v_R = c-v_B$ and $v_L = c + v_B$. The function $u(x',t')$ and
its partial time derivatives are continuous, but its partial space
derivative is not. Integrating Eq. (\ref{eq:mirrorweq}) between $0^-$
and $0^+$ yields the remaining boundary condition. Matching the
solutions on the two sides of the barrier using the boundary
conditions yields
\begin{subequations}
\begin{eqnarray}
\label{eq:M-}
M_{+}(\omega)\, \left( \begin{array}{c} A_\omega^+
  \\ B_\omega^+ \end{array} \right) &=& 
M_{-}(\omega)\, \left( \begin{array}{c} A_\omega^-
  \\ B_\omega^- \end{array} \right)
\end{eqnarray}
where
\begin{equation}
\label{eq:Mpm}
M_{\pm}(\omega) = \left( \begin{array}{cc} 1&1\\ -i\omega
  \,v_L\mp\lambda\, c/2& i\omega \,v_R\mp\lambda\, c/2
\end{array} \right)
\;.
\end{equation}
\end{subequations}
Using Eqs. (\ref{eq:M-}) and (\ref{eq:Mpm}), the scattering matrix
connecting incoming and outgoing amplitudes can be computed:
\begin{equation}
S(\omega) = \left( \begin{array}{cc} \frac{1}{1+i2\,\omega
    /\lambda}&\frac{i2\,\omega /\lambda}{1+i2\,\omega
    /\lambda}\\ \frac{-i2\,\omega /\lambda}{1+i2\,\omega
    /\lambda}&\frac{-1}{1+i2\,\omega /\lambda}
\end{array} \right).
\end{equation}

Now, to determine the heat transmission and reflection coefficients,
one needs to go back to the reference frame of laboratory (i.e., that
of the reservoirs), where the Bose-Einstein occupation numbers of the
phonons are known: $A_\omega^\pm = \left( \frac{c}{v_R} \right)\,
a^\pm_{\omega c/v_R}$ and $B_\omega^\pm = \left( \frac{c}{v_L}
\right)\, b^\pm_{\omega c/v_L}$, where $\langle {a^-_\omega}^\dagger
a^-_\omega \rangle = n_L(\omega)$, $\langle {b^+_\omega}^\dagger
b^+_\omega \rangle = n_R(\omega)$, and $\langle {a^-_\omega}^\dagger
{b^+_\omega} \rangle = \langle {b^+_\omega}^\dagger {a^-_\omega}
\rangle = 0$, since phonons coming from different reservoirs are
uncorrelated. Thus, the heat current leaving the left reservoir is
given by the expression
\begin{equation}
{\cal J}^Q_L = \int_0^\infty d\omega\, \omega \left[ n_L(\omega) -
  \langle {b^-_\omega}^\dagger b^-_\omega \rangle \right].
\end{equation}
The quantity $\langle {b^-_\omega}^\dagger b^-_\omega \rangle$ can be
expressed in terms of the distributions $n_{L,R}(\omega)$ through the
scattering matrix $S(\omega)$. After a few manipulations, we arrive at
\begin{eqnarray}
{\cal J}_L^Q & = & \int d\omega\, \omega \left|
S_{12}\left(\omega\right) \right|^2 \left[ n_L(\omega)- n_R(\omega)
  \right] \nonumber\\ & + & \int d\omega\, \omega\, |S_{11}(\omega)|^2
\left\{\left[ n_L(\omega) - \left( \frac{c}{v_R} \right)^2 n_L \left(
  \frac{\omega c}{v_R}\right) \right] \right. \nonumber \\ & - &
\left. \left[n_R(\omega)-\left( \frac{c}{v_L} \right)^2 n_R \left(
  \frac{\omega c}{v_L} \right) \right]\right\} \;.
\label{eq:I_L}
\end{eqnarray}
The first line of Eq.~(\ref{eq:I_L}) is the thermal heat current
$I_{\rm thermal}$ from left to right in the presence of a nonmoving
barrier. The second line, which we name $I_{\rm pump}$, results from
the barrier motion and it is clearly zero when $v_B\to 0$
($v_L=v_R=c$). In the limit when the barrier amplitude is high,
$\lambda \gg T_{R,L}$, we obtain
\begin{equation}
{\cal J}_L^Q \approx \frac{4\pi^4}{15}\, \frac{1}{\lambda^2 c^2}\,
\left(T_L^4\,v_R^2 - T_R^4\,v_L^2\right).
\label{eq:I_L2}
\end{equation}
Notice that this current is always negative if $T_L<T_R$ and $v_B>0$
(with $v_R>v_L$), thus, as expected, we are fighting this heat flux
with the energy flux of Eq.~(\ref{eq:J-gas}). A net flux of energy is
indeed possible if we satisfy ${\cal J}_L^E + {\cal J}_L^Q > 0$, which
requires
\begin{eqnarray}
\label{eq:TL}
T_L^2 & > & \frac{4\pi^3}{5} \frac{1}{\lambda^2 c v_B} \left( T_R^4
v_L^2 - T_L^4 v_R^2 \right).
\end{eqnarray}
As mentioned earlier, for a fully reflective barrier
($\lambda\rightarrow \infty$), cooling can be obtained for any
temperature gradient. For a semireflective barrier, to leading order
in $v_B/c$, cooling requires $T_L > T_{\rm min}$, where $T_{\rm min}$
is given by Eq. (\ref{eq:Tmin}) with $T_H = T_R$. Notice that when
$T_L = T_R = T$, the proposed mechanism also allows one to transfer
heat between reservoirs provided that $T/\lambda <
(1/2\pi)\sqrt{5/2\pi}$, independently of the barrier speed.

A few remarks are in order. First, we note that the inequality
(\ref{eq:TL}) is independent on $\gamma$. In fact, anharmonicity is
not essential for the operation of the cooling mechanism. Although
anharmonicity is necessary for equilibration to occur in a {\it
  closed} system, it is not so in an {\it open} system coupled to
thermal reservoirs. For the latter, equilibration and thermalization
takes place over time scales of the order of the time required for
sound waves to propagate back and forth through the
system. Straightforward numerical simulations of a harmonic linear
chain of masses and springs coupled to a thermal reservoir at finite
temperature show that, for practical purposes, fast equilibration is
achieved when the barrier moves between reservoirs with a speed a few
times smaller than $c$. This is important because anharmonic effects
are very weak at low temperatures \cite{klemens67,wolfe98} and should
not significantly contribute to equilibration. Second, work is
inevitably done when the barriers are activated and deactivated during
the $B\to C$ and $C\to A^\prime$ processes. However, during a fixed
cycle, this work does not scale with the length of the cavity
connecting the two reservoirs, while the amount of energy extracted
from the cold reservoirs does. Therefore, the contribution of this
work to the energy balance of the cooling process can be made very
small for a sufficiently long cavity and due to this reason we
neglected it in our estimates of the minimum cooling temperature
$T_{\rm min}$. Finally, although Eq. (\ref{eq:I_L}) has been derived
assuming coherent heat transport, Eq. (\ref{eq:J-gas}) does not rely
on quantum coherence. Hence, coherence is not an essential ingredient
for our heat pump.

Finally, let us discuss practical implementations. To produce a
propagating barrier, it is better to use electromechanical couplings
rather than purely mechanical ones, since electronic controlled is
both more precise and allows for faster switching times. Strongly
electrostrictive materials, in which changes in phonon dispersion are
caused by an external electric field, could be used. In particular,
electrostrictive polymers such as poly-vinylidene fluoride (PVDF), in
which giant electrostriction has been observed \cite{zhang98}, appear
to be a promising class of materials for building phonon pumps. Like
other one-dimensional systems, a single chain of PVDF has four
acoustic phonon branches: one longitudinal, two transverse, and one
twist mode. Being a highly ionic (or polar) polymer, PVDF has a
permanent dipole moment per monomer unit which couples to the external
electric field, leading to a gap in the acoustic twist mode
dispersion. Therefore a local electric field can virtually block the
torsion modes with frequencies below the gap from propagating in PVDF
\cite{menezes10}, which is equivalent to introducing an infinite
barrier for such phonons in our scheme.

In this particular implementation, we can understand more clearly
other aspects of the phonon pump. For example, the insertion or
removal of the phonon barrier corresponds to turning on or off the
electric fields. Because the field causes a phonon gap for the
torsional modes, if the insertion is adiabatic, the energy required to
do so is given by the phonon energy density that is excluded from the
barrier region. As long as the barrier is much narrower than the
length of the channel, this energy can be much smaller than the energy
pumped as the barrier is pushed along the polymer. In this case, the
approximation of neglecting the switching on or off of the barrier
(via electric field) holds well. As explained in Ref.
\cite{menezes10}, for an electric field of 10 MV/cm (a typical field
for nanoscale field effect devices), the threshold gap frequency for
PVDF corresponds to a temperature of roughly 5 K. Therefore, if the
device operates at temperatures below this range, these phonons will
be effectively blocked from participating in heat
transmission. Coupling PVDF to a grid of backgate electrodes that can
be individually controled would effectively produce a moving large
barrier potential, as required by our pump. To evaluate the cooling
capability of the pump, let us use 5 K as an estimate for $\lambda$
(set by the threshold gap mentioned above) and a velocity ratio $v_B/c
=1/10$. Then, $\Theta_B\approx 0.4$ K; it follows from
Eq.~(\ref{eq:Tmin}) that if $T_H = 100$ mK, $T_{\rm min} \approx 20$
mK, while if $T_H = 5$ mK, $T_{\rm min} \approx 40$ $\mu$K.

This work is supported in part by the DOE Grant DE-FG02-06ER46316
(CC), CONICET and ANPCyT in Argentina (LA), and CAPES, CNPq, FAPERJ,
and INCT-Nanomateriais de Carbono in Brazil (RBC).

%%%%%%%%%%%%%%%%%%%%%%%%%%%%%%%%%%%%%%%%%%%%%%%%%%%%%%%%%%%%%%%%%%%%%%%%
%%%%

%%%%%%%%%%%%%%%%%%%%%%%%%%%%%%%%%%%%%%%%%%%%%%%%%%%%%%%%%%%%%%%%%%%%%%%%
%%%%

\end{document}